# Optimising MFCC parameters for the automatic detection of respiratory diseases

Yuyang Yan[1,*], Sami O. Simons[2], Loes van Bemmel[2], Lauren Reinders[2], Frits M.E. Franssen[2], and Visara Urovi[1]

[1]Institute of Data Science, Maastricht University, Paul-Henri Spaaklaan 1, Maastricht, 6229 EN, The Netherlands

[2]Department of Respiratory Medicine, Maastricht University Medical Centre, P. Debyelaan 25, 6229 HX Maastricht, The Netherlands

**Abstract**—Voice signals originating from the respiratory tract are utilized as valuable acoustic biomarkers for the diagnosis and assessment of respiratory diseases. Among the employed acoustic features, Mel Frequency Cepstral Coefficients (MFCC) is widely used for automatic analysis, with MFCC extraction commonly relying on default parameters. However, no comprehensive study has systematically investigated the impact of MFCC extraction parameters on respiratory disease diagnosis. In this study, we address this gap by examining the effects of key parameters, namely the number of coefficients, frame length, and hop length between frames, on respiratory condition examination. Our investigation uses four datasets: the Cambridge COVID-19 Sound database, the Coswara dataset, the Saarbrücken Voice Disorders (SVD) database, and a TACTICAS dataset. The Support Vector Machine (SVM) is employed as the classifier, given its widespread adoption and efficacy. Our findings indicate that the accuracy of MFCC decreases as hop length increases, and the optimal number of coefficients is observed to be approximately 30. The performance of MFCC varies with frame length across the datasets: for the COVID-19 datasets (Cambridge COVID-19 Sound database and Coswara dataset), performance declines with longer frame lengths, while for the SVD dataset, performance improves with increasing frame length (from 50 ms to 500 ms). Furthermore, we investigate the optimized combination of these parameters and observe substantial enhancements in accuracy. Compared to the worst combination, the SVM model achieves an accuracy of 81.1%, 80.6%, and 71.7%, with improvements of 19.6%, 16.10%, and 14.90% for the Cambridge COVID-19 Sound database, the Coswara dataset, and the SVD dataset respectively. To validate the generalization of these findings, we employ the Long Short-Term Memory (LSTM) model as a validation model. Remarkably, the LSTM model also demonstrates improved accuracy of 14.12%, 10.10%, and 6.68% across the datasets when utilizing the optimal combination of parameters. The optimal parameters are validated using an external voice pathology dataset (TACTICAS dataset). The results demonstrate the generalization capabilities of the optimized parameters across various pathologies, machine-learning models, and languages.

✦

## 1 INTRODUCTION

RESPIRATORY disease is the third leading cause of death globally, with an annual toll of over 4 million lives lost [1]. The high prevalence of respiratory diseases leads to high medical costs and poses a heavy health burden worldwide.

Traditionally, auscultation has been used for the detection of respiratory disorders. Auscultation involves listening to the sounds produced by the heart and lungs using a stethoscope. However, human contact is essential which may raise concerns about the potential spread of diseases, and a human interpretation of the sounds can introduce variability. Speech signals generated by organs involved in respiratory functions are closely linked to respiratory conditions [2]. Conditions such as chronic obstructive pulmonary disease (COPD), asthma, and COVID-19 significantly influence the vocal tract and result in abnormal respiratory sounds. The exceptional respiratory sounds have shown promise as indicators of underlying respiratory conditions [3] [4], and these signals can be analyzed remotely which makes it less observer-dependent, which opens the opportunity for remote diagnostics (as opposed to the human-to-human interaction in auscultation).

The widespread availability of smart devices has enabled remote healthcare diagnosis, tracking, and control of infectious diseases [5]. Data-driven approaches [6], Computer Aided Detection (CAD) [7] and Artificial Intelligence (AI)-based tools [8] are gaining interest in the field of respiratory disease diagnosis. Particularly, due to the potential risks associated with respiratory diseases, acoustic changes can be identified in respiratory sounds. One representative example is ResApp [9], an application created in 2014, that has shown over 90% accuracy in detecting respiratory diseases using only a smartphone instead of optical imaging (X-ray, CT), blood, and/or sputum tests, which are time-consuming and expensive. Moreover, the ongoing development of COVID-19 screening applications [10] [11] further underscores the continued utilization of smart devices for remote disease diagnosis during outbreaks and future epidemics.

Behind those applications, acoustic feature extraction and machine learning algorithms play an important role in the diagnosis of respiratory diseases. Two primary approaches have emerged in the development of diagnosis systems: traditional pipeline systems and end-to-end systems. Traditional pipeline systems follow a two-step process that involves hand-crafted feature extraction followed by classification. In the feature extraction step, acoustic information is captured using jitter, shimmer, fundamental frequency, Noise to Harmonic Ratio (NHR), and mel-frequency cepstral coefficients (MFCC) [12] [13]. In the second step, machine learning models are trained to identify the pathological cases based on these extracted features. In end-to-end systems, raw speech signals are used as inputs, the features are automatically learned from complex models [14]. However, this approach lacks interpretability and demands a substantial amount of data for effective training. As a result, the traditional pipeline system remains an effective approach for respiratory disease diagnosis and will be utilized in this study.

MFCC has emerged as a popular method for extracting acoustic features from voice signals [15], which represent audio based on the perception of human auditory systems. Initially proposed by Davis and Mermelstein in the 1980s [16], it has been utilized in various audio recognition tasks, including speech emotion recognition [17], speaker recognition [18] and pathology voice identification [19] in traditional pipeline systems. Although MFCC is widely used, the impact of its parameters such as frame length, hop length between frames, and the number of coefficients on MFCC extraction has not been systematically investigated [20]. Tirronen's study [20] examined the impact of the frame length on voice pathology detection. However, the influence of the hop length between frames and the number of coefficients was not addressed. Additionally, the investigation is limited to a single voice pathology dataset, with no exploration into other pathologies or additional non-German speech datasets.

The novelty of this work lies in the systematic exploration of various frame lengths, hop lengths between frames, and numbers of coefficients, aiming to enhance the quality of MFCC for respiratory condition examination. Furthermore, our investigation extends beyond a single voice pathology dataset, incorporating multiple datasets representing diverse vocal pathologies. We employ an external machine learning model to comprehensively assess the impact of MFCC parameters. Utilizing a validation dataset, we confirm that the optimal MFCC parameters result in enhanced accuracy when identifying the presence or absence of a specific respiratory condition. The findings demonstrate the generalization capabilities of the optimized parameters across different pathologies, machine-learning models, and languages.

The paper is structured as follows: Section 2 describes the background of MFCC extraction and the related datasets used in this study. Section 3 outlines the methods, including pre-processing, feature extraction, and classifier description. The detailed results of the experiments are presented in Section 4, followed by a discussion of the results in Section 5. Section 6 provides a summary of the findings. Finally, the limitations of this study and the future directions are identified in Section 7.

## 2 BACKGROUND

### 2.1 Mel Frequency Cepstral Coefficients (MFCC)

The speech generated is influenced by the structure of the vocal tract, it manifests itself in the envelope of short-time power spectrum. The task of MFCC is to capture and represent this envelope appropriately [21].

The MFCC was used to extract features in this study, the specific calculation process is shown in Figure 1, and the detailed procedure of MFCC extraction is given below.

*Step 1. Framing.*

Due to the dynamic nature of the voice signal, it can only be regarded as stable over short durations [22]. As a result, the original time-domain signal is divided into short frames with $N$ samples and shifted by $M$ samples ($M < N$). The framing procedure is illustrated in Figure 2.

The parameter $N$ represents the frame length. Choosing a longer frame length enhances frequency resolution but may introduce frequent changes in information within each frame. Conversely, a shorter frame length optimizes time-domain representation but may not capture sufficient reliable information [23]. $M$ signifies the hop length between frames, where a smaller hop length increases overlap between frames, and a larger hop length results in less overlap. Therefore, optimizing both frame length and hop length is crucial for achieving an effective balance between time and frequency considerations.

*Step 2. Windowing.*

To eliminate discontinuities at the edges of each frame and reduce spectral leakage, a Hamming window is applied.

$$\widetilde{s}(n) = s(n)w(n), 0 \leq n \leq N-1 \quad (1)$$

In (1), $s(n)$ is the original audio frame, $\widetilde{s}(n)$ is the processed voice signal, and $w(n)$ is the Hamming window in (2).

$$w(n) = 0.54 - 0.46\cos\left(\frac{2\pi n}{N-1}\right), 0 \leq n \leq N-1 \quad (2)$$

*Step 3. Fast Fourier Transform (FFT).*

Then implement FFT to each frame, where $i$ indicates the number of the frame.

$$S_i(k) = \sum_{n=1}^{N} s_i(n)w(n)e^{-j2\pi kn/N}, 1 \leq k \leq K \quad (3)$$

Where $S_i(k)$ is the FFT of each frame and K is the length of FFT.

*Step 4. Band-pass filtering.*

A set of triangular band-pass filters is applied to the spectrum.

$$E_i = \sum_{k=0}^{N/2-1} \phi_j(k)A_k, 0 \leq j \leq J-1 \quad (4)$$

Where $\phi_j$ is the $j$th filter, the total number of filters is J, $A_k$ is the amplitude of $S_i(k)$.

$$A_k = |X(k)|^2, 0 \leq k \leq N/2 \quad (5)$$

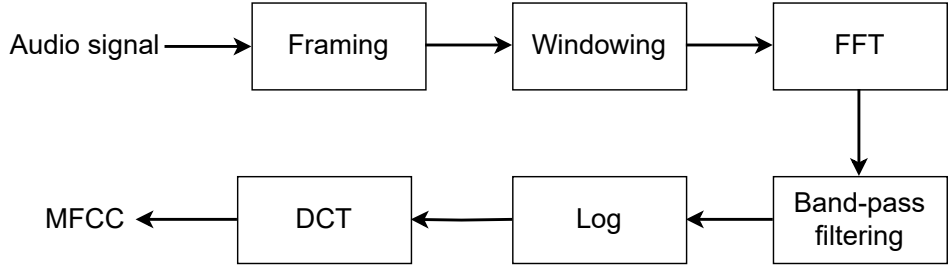

Fig. 1: The overall process of MFCC

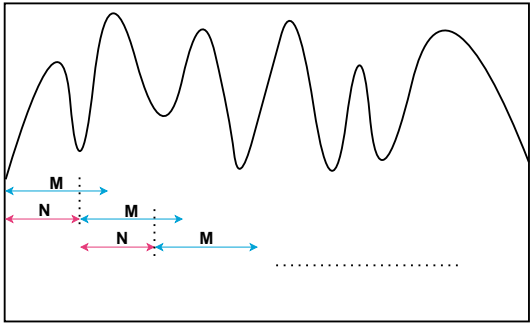

Fig. 2: Audio framing illustration

*Step 5. Logarithm (log).*

Take the logarithm of all filterbank energies to separate the excitation spectrum from the vocal system spectrum.

*Step 6. Discrete Cosine Transform (DCT).*

The Basic concept of DCT is to correlate the value of the spectrum to produce essential information about the signal structure [24].

$$C_m = \sum_{j=0}^{J-1} \cos(m\frac{\pi}{J}(j+0.5)) log_{10}(E_j), 0 \leq m \leq L-1 \quad (6)$$

Finally, the number of coefficients is $L$. The low-order cepstral coefficients exhibit characteristics that are sensitive to the spectral slope, while the higher-order coefficients are more sensitive to noise. Consequently, only the first few cepstral coefficients are selected, as the higher order coefficients represent the excitation process, which is less informative for accurately depicting the shape of the power spectrum [25]. Therefore, optimizing the number of coefficients becomes crucial for achieving an effective balance between capturing relevant spectral information and minimizing sensitivity to noise.

## 2.2 Datasets

This study uses four datasets: the Cambridge COVID-19 Sound database, the Coswara dataset, the Saarbrücken Voice Disorders (SVD) database, and the TACTICAS dataset, which are summarized as follows. Further details regarding the dataset characteristics can be found in Table 1.

The Cambridge COVID-19 Sound database was collected through a mobile and web application [19]. Participants were required to provide their medical history, smoking habits, and other general demographic information. They could then upload their sounds along with their COVID-19 status. The audio recordings consist of three voluntary cough sounds, three to five breathing sounds, and speech recordings where the user was asked to read a specific sentence ("I believe my data may assist to manage the virus pandemic"), most of the users read in English. In this study, we only focus on the speech signals within the datasets. After a year of collection, a total of 893 speech samples were released.

The Coswara dataset was also collected for the COVID-19 study through a web and mobile application [26]. It includes nine sound categories such as breathing (shallow and deep), cough (shallow and heavy), sustained vowel phonation (/a/, /e/, and /o/), and 1 - 20 counting (normal and fast speed) in English. Additional metadata information, including age, gender, location (country and state), current health status (healthy, exposed, cured, or infected), and the presence of comorbidity (pre-existing medical conditions) was also collected. In our analysis, we specifically focus on the speech signals within the Cambridge COVID-19 Sound database, thus we also utilize only speech recordings (normal and fast speed counting) from the Coswara dataset. As indicated in Table 1, there are 1350 positive cases and 3898 negative speech cases in the Coswara dataset.

The Saarbrücken Voice Disorders (SVD) database is an open-access database that contains audio recordings from more than 2000 individuals, mostly native German speakers [27]. There are 71 pathologies in the SVD database, we specifically select three disorders (hyperkinetic dysphonia, hypokinetic dysphonia, and reflux laryngitis) from the SVD database in this study. We also use only speech recordings of the German sentence "Guten Morgen, wie geht es Ihnen" ("Good morning, how are you?") in the SVD dataset. There are 355 positive participants for the three selected disorders and 588 healthy speakers in this dataset.

The TACTICAS [28] dataset was collected through a mobile application. On the first day of data collection, the participants go to the hospital and report their population parameters including age, gender, height, and weight. Their baseline information including clinical characterization, medical examination, lung function, and symptoms was also collected. The participants were previously diagnosed with asthma or Chronic obstructive pulmonary disease (COPD). In the following days, the participants were asked to upload their sustained vowel /a/, answering of an open question, reading of a short story and fill out the E-RS (EXACT-Respiratory Symptoms) questionnaire daily through a mobile application, all three sound events were used in this study as shown in Table 1. We labeled the data based on RS-Breathlessness, which refers to the component of the E-RS questionnaire that assesses the severity and impact of breathlessness or shortness of breath experienced by individuals with respiratory conditions. The meaningful threshold for breathlessness symptom decline is 1 point within each subject [29].

## 3 METHODS

### 3.1 Pre-processing

Speech recordings in different datasets have varying lengths. The average speech duration in the Cambridge COVID-19 Sound database is 13.1 seconds, in the Coswara dataset and the TACTICAS dataset is 11.1 seconds and 14.7 seconds respectively, while in the SVD dataset, it is 2 seconds. To ensure a consistent analysis, we initially divided the audio recordings from the Cambridge COVID-19 Sound database, the Coswara dataset, and the TACTICAS dataset into 3-second segments. The silent portions within the recordings were identified by a specified decibel threshold (mean value minus the standard deviation), which was slightly below the average loudness of the audio signal, considering the standard deviation [30]. Given the focal point of our analysis on individual segments, we remove the silent portions at the segment level. Subsequently, any intervals with sound levels below this threshold were eliminated. We only retained the remaining recordings that are longer than 1 second, discarding the shorter ones. Following the pre-processing steps, the audio lengths were approximately equal, with an average speech duration of 2.3 seconds in the Cambridge COVID-19 Sound database, 2.2 seconds in the Coswara dataset, 2.4 seconds in the TACTICAS dataset and 1.6 seconds in the SVD dataset.

### 3.2 Feature extraction

This study investigated three primary parameters: the number of coefficients ($L$ in Section 2.1), the frame length ($N$ in Figure 2), and the hop length ($M$ in Figure 2) between frames. The default MFCC parameters include 13 coefficients, a frame length of 25 ms, and a hop length of 10 ms between frames. To extract MFCC, the default set of coefficients is 13, but 20 to 60 coefficients are also widely used [31] [32], thus we employed eight values (13, 20, 30, 40, 50, 60, 70, 80) for the number of coefficients investigation. For the frame length, we explored eight values (25 ms, 50 ms, 100 ms, 200 ms, 300 ms, 400 ms, 500 ms, 800 ms) to compare with Tirronen's study [20], whose investigation is from 25 ms to 500 ms. Furthermore, we employed eight values (5 ms, 25 ms, 50 ms, 100 ms, 200 ms, 300 ms, 400 ms, 500 ms) for the hop length between frames to extract MFCC, since this parameter is usually set from 10 ms to 500 ms in pathology detection [31], [32]. When one parameter was investigated, the other parameters were held constant. The MFCC vectors were obtained by calculating the mean from the frame-wise MFCC values. The Spafe library was utilized to extract MFCC features from each recording [33].### 3.3 Classifiers

A binary classifier was applied across diverse datasets to determine the presence/absence of a specific respiratory condition recorded within each dataset. The binary classifier employed was a Support Vector Machine (SVM) model implemented through the Scikit-learn library [34]. SVM is a versatile algorithm capable of handling both classification and regression problems [35]. It operates by plotting each observation as a point in an n-dimensional space (where n is the number of features in the dataset) and aims to find an optimal hyperplane that can effectively separate the data points into their respective classes. The generalization parameter, C, was set to 1.0, indicating the balance between achieving a low training error and maintaining a wide margin. The kernel function employed was the radial basis function (RBF), which allowed for non-linear separation of data. The gamma value, set to 0.1, determined the influence of each training sample, with higher values indicating a more localized decision boundary. The pipeline of the detection system is shown in Figure 3.

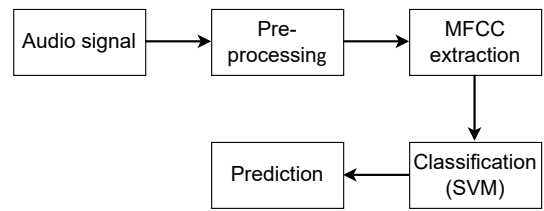

Fig. 3: Schematic representation of the detection system

To assess the performance of the optimized MFCC parameters with other models, the Long short-term memory (LSTM) model was used for validation. The LSTM model is known for its ability to learn long-term dependencies [36] and capture complex patterns in sequential data, making it widely used in voice pathology detection [37] [38]. In this study, the LSTM model was constructed with one Bidirectional LSTM layer, followed by a dropout layer and one dense layer. A sigmoid activation function was employed in the output dense layer for classification. The model was trained using an Adam optimizer, with a maximum epoch of 100 and a batch size of 32. The LSTM model was built based on our previous research [39] on COVID-19 detection, and no parameter fine-tuning was conducted specifically for this study.

To evaluate the performance of the model, all experiments were conducted using 10-fold cross-validation. This technique involved dividing the dataset into ten subsets, where one fold was used for testing while the remaining nine folds were used for training. This process was repeated ten times, with each fold serving as the test set once. The results obtained from these iterations were then analyzed to derive the mean and standard deviation, providing a comprehensive understanding of the models' performance.

To ensure the same ratio of positive and negative cases in each fold, the StratifiedKFold function from the Scikit-learn library [34] was employed. This function ensured that the distribution of positive and negative instances remained consistent across all folds, avoiding potential bias in the evaluation process.

TABLE 1: Patient characteristics in the databases used in this study

| Datasets | Sound Event | Main language | Age | | Disease status (n) | | Gender (n) | | |
|---|---|---|---|---|---|---|---|---|---|
| | | | Min | Max | Positive | Negative | Male | Female | Others |
| Cambridge database | Reading | English | 9.5 | 90 | 308 | 585 | 204 | 157 | 5 |
| Coswara dataset | Counting | English | 15 | 90 | 1324 | 3854 | 1789 | 806 | 2 |
| SVD database | Reading | German | 9 | 94 | 355 | 588 | 380 | 563 | 0 |
| TACTICAS database | Reading, Answering, Vowel | Dutch | 35 | 83 | 2816 | 3528 | 16 | 17 | 1 |

## 4 RESULTS

The accuracy is considered an important indicator in finding the optimized parameters since it represents the percentage of correctly classified cases, making it easy for healthcare professionals, and the general public to understand the performance. To better evaluate the performance of the optimized parameters, we list several indicators used in TACTICAS dataset validation, including accuracy, AUC (Area under the ROC curve), F1 score, precision, and EER (equal error rate).

### 4.1 Number of coefficients

Choosing varying numbers of coefficients indicates the utilization of different quantities of the initial cepstral coefficients. Our investigation covered a range of coefficients from 13 to 80, where 13 coefficients denote the selection of the first 13 coefficients, and 80 coefficients indicate the inclusion of the first 80 coefficients. The other two parameters were set to the default setting: the frame length was set to 25 ms, and the hop length between frames was set to 10 ms.

The obtained results for different numbers of coefficients are presented in Figure 4. In the case of the Cambridge COVID-19 Sound database, the accuracy of the SVM model initially improved as the number of coefficients increased. However, after reaching a peak at around 30 coefficients, the accuracy began to decline. Similar trends were observed for both the Coswara dataset and the SVD dataset, where the highest performance was achieved when the number of coefficients was 30 to 40, followed by a decrease in accuracy.

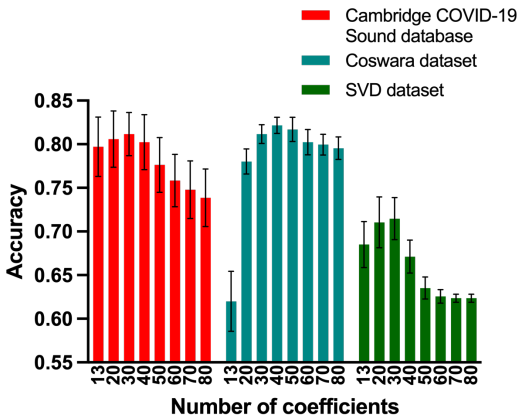

Fig. 4: Different number of coefficients

### 4.2 Frame length

In the procedure of MFCC extraction, the first step involves dividing the origin time-domain signal into short frames, where the duration of each frame is defined as the frame length, denoted as $N$ in Figure 2. This step is essential because the frequencies in a signal change over time, performing a Fourier transform across the entire signal would lead to the loss of frequency contours over time.

The hop length was set as default 10 ms, and based on the results in Section 4.1, the number of coefficients was set as 30. As shown in Figure 5, for the Cambridge COVID-19 Sound database and the Coswara dataset (both of which are COVID-19 pathology), the accuracy of the SVM model decreased with increasing frame length. This implies that the performance of MFCC deteriorated when the original time-domain signal was divided into longer durations. In Figure 5, the highest accuracy for each dataset was achieved at a frame length of 25 ms. However, concerning the SVD database, the highest accuracy was also achieved at 25 ms, but exhibited an overall increasing trend from 50 ms to 500 ms, with a decrease observed at 50 ms, this trend aligns with the findings of prior work on SVD dataset conducted by Tirronen's [20].

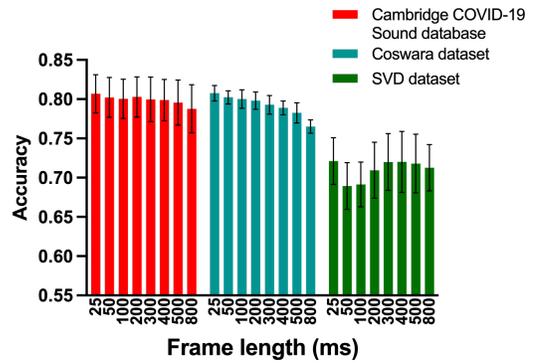

Fig. 5: Different frame lengths

### 4.3 Hop length between the frames

The final parameter explored in this study was the hop length, which refers to the number of samples shifted in each framing step, denoted as $M$ in Figure 2. In this section, the number of coefficients and the frame length were set as 30 and 25 ms which were the optimized findings in Section 4.1 and Section 4.2. Typically, the hop length is smaller than the frame length, allowing for some overlap between frames. In the context of MFCC extraction, a window function is employed to minimize spectral leakage. However, this process also disregards the signal at window boundaries. Overlapping frames can help compensate for this loss, providing a more comprehensive representation of the audio signal. As depicted in Figure 6, the performance of MFCC decreased as the hop length increased. The highest accuracy for each dataset was achieved with a hop length of 5 ms, suggesting an overarching decreasing trend as more samples are shifted in each framing step.

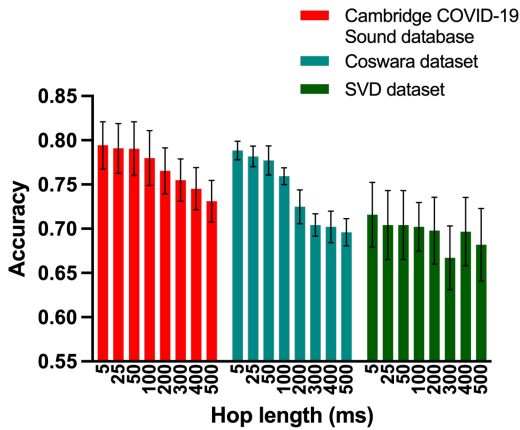

Fig. 6: Different hop lengths

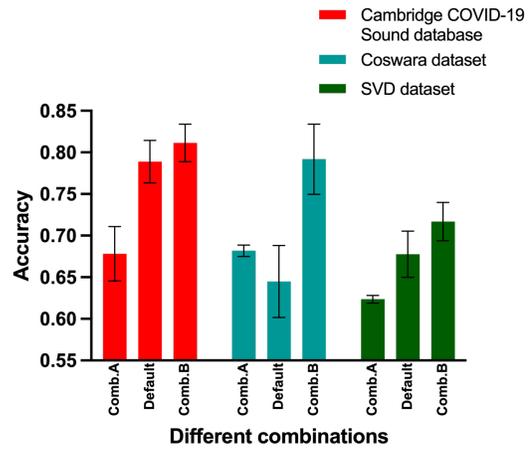

Fig. 7: Different combinations with the SVM model

### 4.4 Optimized combination of parameters

To optimize the performance of MFCC, the optimized parameter combinations were investigated. Based on the previous sections, for the Cambridge COVID-19 Sound database and the Coswara dataset, the highest accuracy was achieved with a shorter frame length and a shorter hop length, and the number of coefficients around 30. For the SVD database, the highest accuracy was also achieved with a shorter hop length and the number of coefficients around 30, but for the frame length, there were several candidates, except 50 ms and 100 ms. Considering these findings, the optimized parameters combination (Comb. B in Figure 7) was defined as a frame length of 25 ms, a hop length of 5 ms, and 30 coefficients. Conversely, the worst parameters combination (Comb. A in Figure 7) comprised a frame length of 800 ms, a hop length of 500 ms, and 80 coefficients. In addition, the default combination of parameters (Default in Figure 7) was also investigated, the frame length was set to 25 ms, the hop length was set to 10 ms, and the number of coefficients was set to 13 by default [40]. The optimized parameters combination (Comb. B in Figure 7) achieved the highest accuracy compared with the default and worst combination. The mean accuracy improved by 19.6%, 16.10%, and 14.90% compared with the worst combination, and improved by 2.78%, 22.79%, and 5.75% compared with the default parameters for the Cambridge COVID-19 Sound database, the Coswara dataset, and the SVD dataset, the highest accuracy achieved by the optimized parameters were 81.1%, 80.6% and 71.7%, respectively.

### 4.5 Optimized parameters performance for different genders

The performance of the optimized parameters for different genders was also investigated. We respectively split the three datasets into two groups of gender: Male and Female (the 'Other' group in Table 1 was discarded), and trained the model again. For each gender, we compared the performance of different combinations, all the combinations of parameters remained consistent with those defined in the previous section 4.4. The results from Figure 8 indicate the accuracy improved for both male and female groups, the improvement was more in the Cambridge COVID-19 Sound database and the Coswara dataset. For the male group in the SVD dataset, the accuracy slightly decreased compared with the default parameters.

### 4.6 Long short-term memory (LSTM) model validation

The optimized combination of parameters identified in the previous Section 4.4 was determined based on the SVM model. For the LSTM model validation, both Comb. A and Comb. B parameter combinations remained consistent with those defined in the previous section 4.4. Once again, the optimized parameters combination (Comb. B in Figure 9) achieved the highest accuracy when compared to the default and worst combinations. The accuracies of the LSTM model were 79.2%, 79.6% and 71.9%, which improved by 14.12%, 10.10%, and 6.68% compared with the worst combination, and improved by 6.59%, 3.11%, and 2.57% compared with the default parameters for the Cambridge COVID-19 Sound database, the Coswara dataset, and the SVD dataset, respectively.

### 4.7 External dataset validation

To validate the optimized parameters in the other vocal pathology dataset, we used the TACTICAS dataset. The optimized and the worst parameter combinations remained consistent with those defined in the previous section 4.4. Both the SVM and LSTM model shows the optimized parameters achieved the best performace in Table 2, which accuracies improved by 7.81%, and 5.08% compared with the worst combination, and improved by 14.65%, and 5.22% compared with the default parameters for the SVM and LSTM model.

## 5 DISCUSSION

This study shows that the performance of MFCC can be improved by optimizing the parameters in MFCC extraction. We proposed that the optimization of these parameters should be included in the analysis of respiratory sounds, as it has the potential to enhance prediction accuracy. In Section 4, we observed improvements of up to 19.6%.

TABLE 2: TACTICAS dataset validation

|  | SVM model | | | | | LSTM model | | | | |
| --- | --- | --- | --- | --- | --- | --- | --- | --- | --- | --- |
| Combination | Accuracy | AUC | F1 | Precision | EER | Accuracy | AUC | F1 | Precision | EER |
| Worst | 0.704 | 0.689 | 0.621 | 0.710 | 0.186 | 0.768 | 0.763 | 0.732 | 0.739 | 0.199 |
| Default | 0.662 | 0.666 | 0.643 | 0.603 | 0.372 | 0.790 | 0.784 | 0.754 | 0.775 | 0.166 |
| Optimized | **0.759** | **0.751** | **0.709** | **0.751** | **0.185** | **0.806** | **0.802** | **0.776** | **0.785** | **0.163** |

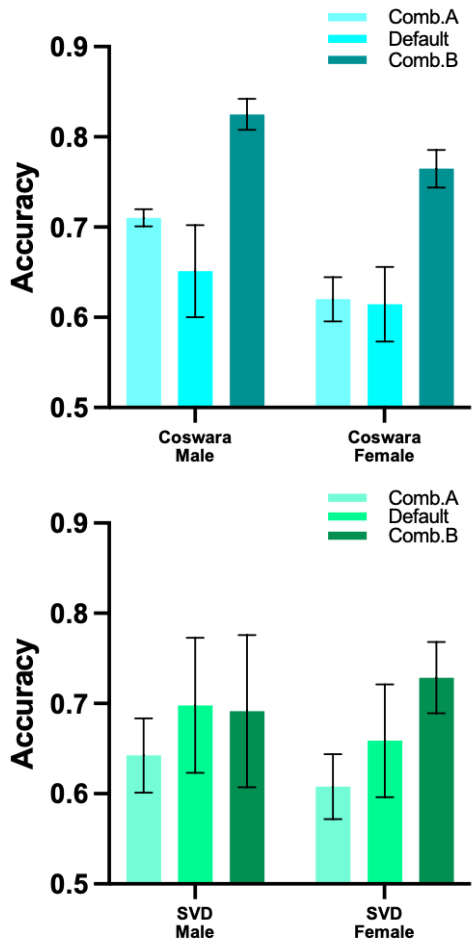

Fig. 8: Different combinations for different genders

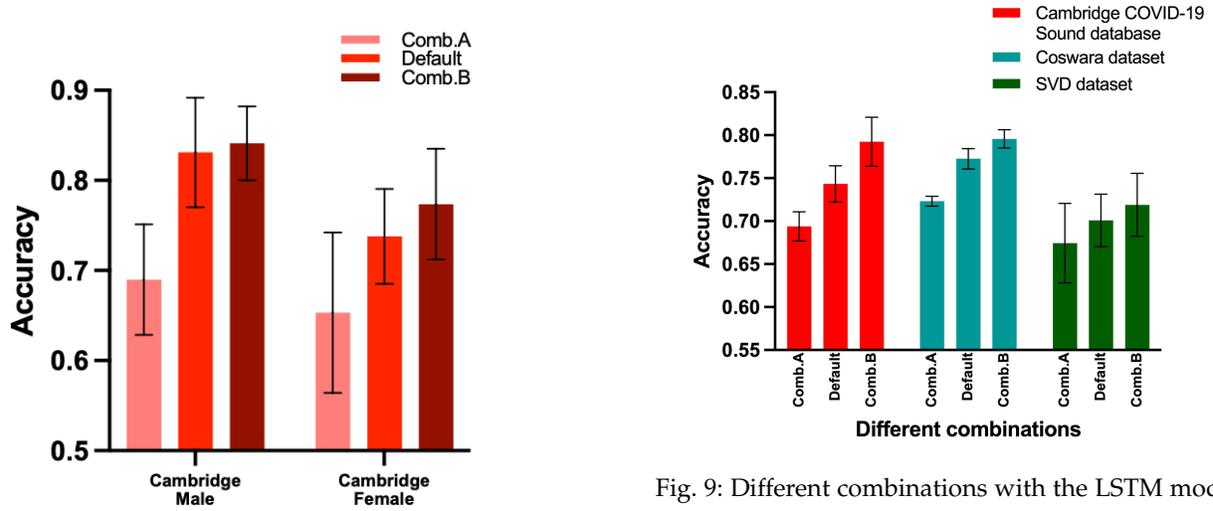

Fig. 9: Different combinations with the LSTM model

The three datasets exhibited similar trends with regard to the number of coefficients. For the Cambridge COVID-19 Sound database, the performance of MFCC improved as the number of coefficients increased, reaching peak accuracies at 30 coefficients. Beyond this point, however, the performance of MFCC started to decline. This observation can be explained by the nature of MFCC extraction, where the first coefficient represents the average power in the spectrum, and lower-order coefficients describe the overall spectral envelope of the speech recordings. Higher-order coefficients, on the other hand, capture finer spectral details such as pitch and tonal information [41], which typically do not contribute significantly to the diagnosis of respiratory diseases. Similar trends were also observed in the Coswara dataset and the SVD database. In the Coswara dataset, the accuracy improved with the number of coefficients until reaching 40, after which it began to decrease. In the SVD database, the accuracy increased with the number of coefficients until approximately 30, beyond which it started to decline. The findings suggest that optimal performance of MFCC is achieved when the number of coefficients is set around 30 for respiratory condition determination.

The impact of frame length was also investigated, revealing varying effects on the performance of the COVID-19 datasets (the Cambridge COVID-19 Sound database and the Coswara dataset) and the SVD database. The prediction accuracy of the COVID-19 datasets shows a decrease as the frame length increases. However, in the case of the SVD database, The highest accuracy was achieved at a frame length of 25 ms. Afterward, the accuracy dropped with frame length (50 - 200 ms) and improved again with higher frame length (300 - 500 ms). In practiced terms, the frame length should not be too short such that we can obtain reliable spectral estimates for each frame. At the same time, it should not be too long such that under a particular frame, the voice sample is time-invariant. Although the audio signal is nonlinear and time-varying, it exhibits steadiness within short time intervals. Consequently, short-time characteristics can typically be extracted by employing a frame length of 10 ms to 30 ms. However, the 50 ms frame length yielded the lowest accuracy in Figure 5 for the SVD dataset. This can be attributed to the nature of the three disorders in the SVD dataset: hyperkinetic dysphonia, hypokinetic dysphonia, and reflux laryngitis, all of which are laryngeal [20] disorders. In the case of these disorders, capturing precise vocal tract information for each phoneme (which is known to be one of the advantages of using MFCCs) may not be as crucial as it is for the other two COVID-19 datasets.

The final parameter investigated in this study was the hop length between frames, the performance of MFCC decreased as increasing hop length. Notably, the highest accuracy was achieved with a 5 ms hop length. A longer hop length for a given frame length results in a smaller overlap between frames and consequently fewer frames overall. Generally, a greater overlap provides more analysis points, which can help avoid or smooth out discontinuities as each segment exhibits a seamless transition to the next. Additionally, a smaller hop length can effectively reduce spectral leakage by compensating for energy loss through the utilization of a window function.

Furthermore, we examined the optimal combination of the three investigated parameters. The mean accuracy demonstrated notable improvements of 19.6%, 16.10%, and 14.90% compared with the worst combination, and improved by 2.78%, 22.79%, and 5.75% compared with the default parameters for the Cambridge COVID-19 Sound database, the Coswara dataset, and the SVD dataset, respectively. These findings highlighted the influence of the investigated parameters on the performance of MFCC, underscoring the importance of selecting the best combination of these parameters to achieve substantial classification optimization.

The optimized combination of parameters, as determined by the SVM model, was further validated using an LSTM model to assess its performance. The results showed that the accuracy of the LSTM model improved by 14.12%, 10.10%, and 6.68% compared with the worst combination, and improved by 6.59%, 3.11%, and 2.57% compared with the default parameters for the Cambridge COVID-19 Sound database, the Coswara dataset, and the SVD dataset, respectively. This indicates that the combination of MFCC parameters not only significantly influences the performance of the SVM model but also exhibits generalization capabilities across different models.

To verify the performance of the optimized parameters, we extracted MFCC again with a new pathology dataset (the TACTICAS dataset), and trained MFCC with both SVM and LSTM models. Again, the optimized parameters achieved the best performance in all indicators in Table 2 for both SVM and LSTM models, with accuracy improved by 7.81%, and 5.08% compared with the worst combination, and improved by 14.65%, and 5.22% compared with the default parameters for SVM and LSTM models, respectively. These results show generalization across different pathologies for our findings.

Moreover, the datasets used in this study vary in languages as mentioned in section 2.2. Specifically, the Cambridge COVID-19 Sound database and the Coswara dataset are primarily in English, the SVD dataset is in German, and the TACTICAS dataset is in Dutch. This diversity underscores the cross-language generalization aspect of our study.

## 6 CONCLUSION

We have discussed the impact of three parameters on MFCC extraction, namely the number of coefficients, frame length, and hop length between frames. We specifically focused on their impact on the respiratory condition examination and aimed to determine the optimal combination of these parameters. Our results revealed that the best combination of parameters led to a substantial improvement in accuracy for different voice pathology datasets. These findings underscore the influence of the investigated parameters on MFCC performance and emphasize the substantial benefits of utilizing the optimal combination of these parameters for respiratory condition examination.

## 7 LIMITATIONS AND FUTURE WORK

One major limitation of this study lies in the limited availability of databases specifically designed for respiratory disorders. We only identified four available databases with speech recordings to support our study. Future research endeavors should prioritize exploring the effects of the investigated parameters across other respiratory disorders. Additionally, our findings indicate that the frame length may have different influences on different respiratory conditions. The clinical reasons for this observation warrant further exploration, as understanding the underlying clinical reasons is essential. In pursuit of clinical applicability, it is crucial to establish the interpretability of acoustic features to understand their meaning and their correlation with associated respiratory conditions in real-world scenarios.

### CONTRIBUTION

The authors confirm their contribution to the paper as follows: Y.Yan was in charge of data analysis and the experiments. Dr. V.Urovi led the research questions, Dr. Md S.O.Simons provided medical feedback, L.Bemmel, L.Reinders and F.M.E. Franssen collected data and provided feedback to the final manuscript. All authors have approved the final manuscript.

### ACKNOWLEDGMENTS

The work was supported by NWO Aspasia grant (no: 91716421). The authors would like to thank the Cambridge University for sharing the data collected via the COVID-19 app, the Indian Institute of Science Bangalore for opening the Coswara dataset, the Saarland University who collected the Saarbrücken Voice Disorders (SVD) database, and the TACTICAS study for the data collection of TACTICAS dataset.